# 1D to 3D Dimensional Crossover in the Superconducting Transition of the Quasi-One-Dimensional Carbide Superconductor Sc$_3$CoC$_4$

Mingquan He[1], Chi Ho Wong[1], Dian Shi[1], Pok Lam Tse[1], Ernst-Wilhelm Scheidt[2], Georg Eickerling[2], Wolfgang Scherer[2], Ping Sheng[1], and Rolf Lortz[1]

[1]Department of Physics and the William Mong Institute of Nano Science and Technology, Hong Kong University of Science and Technology, Clear Water Bay, Kowloon, Hong Kong, China

[2]CPM, Institut für Physik, Universität Augsburg, 86159 Augsburg, Germany

The transition metal carbide superconductor Sc$_3$CoC$_4$ may represent a new benchmark system of quasi-1D superconducting behavior. We investigate the superconducting transition of a high-quality single crystalline sample by electrical transport experiments. Our data show that the superconductor goes through a complex dimensional crossover below the onset $T_c$ of 4.5 K. First, a quasi-1D fluctuating superconducting state with finite resistance forms in the [CoC$_4$]$_\infty$ ribbons which are embedded in a Sc matrix in this material. At lower temperature, the transversal Josephson or proximity coupling of neighboring ribbons establishes a 3D bulk superconducting state. This dimensional crossover is very similar to Tl$_2$Mo$_6$Se$_6$, which for a long time has been regarded as the most appropriate model system of a quasi-1D superconductor. Sc$_3$CoC$_4$ appears to be even more in the 1D limit than Tl$_2$Mo$_6$Se$_6$.

## Introduction

The recent discovery of superconductivity in the transition metal carbide Sc$_3$CoC$_4$ [1,2] brings a new member into the family of quasi-one-dimensional (quasi-1D) superconductors, in line with e.g. the organic Bechgaard salts [3] polysulfur nitride (SN)$_x$ [4], the transition-metal trichalcogenides [5] the quasi-1D relatives of the Chevrel phases such as Tl$_2$Mo$_6$Se$_6$ and In$_2$Mo$_6$Se$_6$ [6,7], arrays of 4 Ångstrom superconducting carbon nanotubes grown in AFI zeolite matrices (CNT@AFI) [8-11] and 5nm thin Pb nanowire arrays synthesized in SBA-15 mesoporous silica [12]. Sc$_3$CoC$_4$ crystallizes in the space group *Immm* and contains one-dimensional [CoC$_4$]$_\infty$ ribbons oriented in the crystallographic *bc* plane and extended along the crystallographic *b* direction. These ribbons are embedded in a scandium matrix and it has previously been shown that the chemical bonding between the carbon atoms of the ribbons and the side-on coordinated Sc atoms is characterized by weak but covalent interactions [13].

The compound undergoes two phase transitions above the superconducting transition with signatures in the specific heat, the electrical resistivity and the magnetic susceptibility [1]. A transition at $T$ = 143 K has been ascribed to a charge-density wave formation. At 72 K, the chains undergo a structural phase transition [1]. This transition has been interpreted as a Peierls-type distortion with alternating out-of-plane displacements of the cobalt atoms above and below the [CoC$_4$]$_\infty$ ribbons [1,14]. The Peierls-transition is driven by the formation of weak Co⋯Co bonding interactions which form quasi-1D chains of cobalt atoms perpendicular to the [CoC$_4$]$_\infty$ ribbons. Accordingly, the quasi-1D [CoC$_4$]$_\infty$ ribbons become *geometrically coupled* via Co⋯Co contacts, but *electronically decoupled* by the opening of a gap in the electronic band structure at the Fermi level [14]. As a third consequence of this Peierls-type distortion we observe a symmetry reduction during the transition from the high-temperature orthorhombic (space group *Immm*) to a low-temperature monoclinic structure (space group *C*2/*m*). In the crystallographic coordinate system of this low-temperature phase, the $[CoC_4]_\infty$ ribbons are oriented perpendicular to the monoclinic *b* direction and the line bisecting the crystallographic *a* and *c* axes [1,2]. Due to the fact that the structural differences are rather small between the high- and the low-temperature phase we will for simplicity reasons refer in the following to the high-temperature crystallographic coordinate system. In polycrystalline samples the resistivity showed a semiconducting increase during the cooling sequences below the Peierls transition temperature, although the resistance remains finite above 4.5 K, below which it becomes superconducting [1,2]. Bulk superconductivity has been probed in these polycrystalline samples in the resistivity, magnetization and specific heat [1,2]. The temperature dependence of these quantities showed a remarkable continuous transition with a nearly linear drop of the resistivity and the volume susceptibility below 4.5 K. Zero resistivity is only established at very low temperatures of ~ 0.5 K. This behavior is typical for strongly quasi-1D superconductors and is associated to the $[CoC_4]_\infty$ ribbons [1,2].

According to the Hohenberg-Mermin theorem [15,16], a long-range phase-coherent superconducting state with zero resistance cannot be established in the 1D limit, unless at absolute zero temperature. A 2D superconductor can overcome this limitation by undergoing the famous Berezinskii-Kosterlitz-Thouless (BKT) transition [17-19]. The BKT transition represents a phase ordering transition triggered by the formation of vortex-antivortex pairs through which a quasi-long-range ordered state is stabilized. For quasi-1D systems with some transversal coupling among parallel 1D chains it has been shown by the mean-field theory that a transition towards a 3D long-range ordered state can occur under certain conditions [20-27].

In $Sc_3CoC_4$ the ribbons form a 3D network of parallel chains with inter-ribbon distances of 3.39/~6.60 Å in the high-temperature phase along the crystallographic *a*/*c* direction, respectively [13]. In the low-temperature phase these distance do not change significantly, *i.e.* the averaged Co$^{..}$Co distance corresponding to the inter-ribbon distance along the *a* direction of the high-temperature structure is 3.38Å [14]. If the superconducting coherence length exceeds the ribbon distance, the ribbons would become weakly transversally coupled via the Josephson or the Proximity effect. Such a coupling could therefore establish a 3D bulk superconducting state in the low-temperature regime. Indeed, we have reported such complex dimensional crossovers in several quasi-1D superconducting systems recently [7,10,12], and the differential resistivity as function of injected current has proven extremely useful for studying such dimensional crossover behaviors. A superconductor in the purely 1D limit exhibits a triangular shaped supercurrent gap, which is shaped by longitudinal phase slips in the order parameter [28,29], whereas a bulk 3D superconductor shows the characteristic textbook rectangular-shaped gap framed by two pronounced peaks at the value of the critical currents and a zero resistance plateau around zero current. In order to investigate in detail the superconducting transition of this extremely low-dimensional superconductor, we present an investigation of the electrical resistivity, *I-V* characteristic and differential resistivity.

**Experimental**

The sample used for this study represented a high quality single crystal needle with dimension 500 μm × 70 μm × 20 μm. Single crystals were grown from a polycrystalline sample of $Sc_3CoC_4$ which was prepared by arc-melting of pure elements (Sc: 3N8; Co: 4N3; C: 5N5) under a highly purified argon atmosphere (500 mbar). The elements were placed together into a water cooled melting pot. Already after the first melting procedure single crystals in form of needles (whiskers) develop spontaneously. To ensure highest possible homogeneity, the polycrystalline sample was turned upside-down and re-melted up to six times. The harvesting of the whiskers was carried out under a microscope using a scalpel or a capillary. Further details on the sample preparation and characterization can be found elsewhere [13].

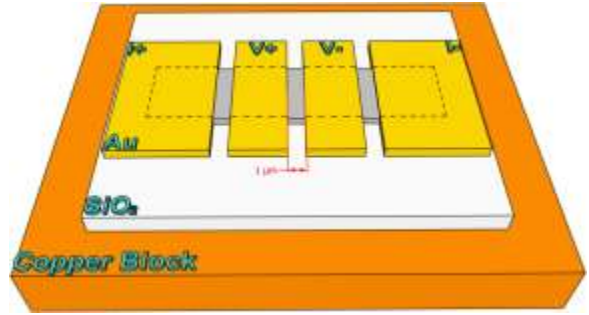

**Figure 1.** Cartoon of our 4-probe device for electric transport measurements on $Sc_3CoC_4$. The gray rectangle represents the sample crystal which is fixed on a $SiO_2$ substrate and partly covered by Au electrodes. The distance between the voltage leads is 1 μm. The $SiO_2$ substrate is attached to the big copper block with good thermal conduction, which serves as a thermal bath to minimize heating effects.

The electrical contacts were created by sputtering a layer of 150 nm Au and subsequently separated by a focused ion beam (FIB, Seiko SMI2050). Figure 1 demonstrates the model of the contact configuration. The separation of the voltage contacts is 1 μm. The temperature dependence of the electrical resistivity was measured with a Keithley 6221 AC/DC current source in combination with a Stanford Research 830 lock-in amplifier. The amplitude and the frequency of the injected current were 50 μA and 5 Hz respectively. We carefully tested the dependence of the resistance as a function of injected current amplitude and found only minor changes in the residual resistance at 2 K in the resistivity for currents smaller than this value. The working frequency was chosen low in order to keep phase shifts of the lock-in signal below 1 degree. The current-voltage measurements at various fixed temperatures were performed with a Keithley 6221 AC/DC current source in combination with a Keithley 2182A nanovoltmeter. The applied current range was -10 mA to 10 mA. In order to minimize heating effects due to the large injected currents, the sample was mounted firmly on a $SiO_2$ substrate, which is in good thermal contact with a large copper block as shown in Figure 1. The copper block served as a thermal bath to absorb the released heat. Furthermore, the current was sent in a short pulse mode. We compared the critical current at fixed temperature of the thermal bath with different pulse time lengths and found that the critical currents were identical if the pulse time length was kept shorter than 0.5 ms. All measurements were thus performed with pulse lengths of 0.1 ms to ensure that no heating effects influence the measurements.

**Results**

The main figure in Figure 2 shows the temperature dependence of the electrical resistivity under various applied magnetic fields. The magnetic field was applied along the crystallographic *a* direction of the sample, which is perpendicular to the $[CoC_4]_\infty$ ribbons. A clear broad superconducting transition is found, which gets gradually suppressed in applied magnetic fields.

The broadening of the transition is a characteristic signature for the presence of phase fluctuations, related to the low-dimensional nature of the system. The resistivity at zero field was measured down to 600 mK and zero resistivity is only established below 1.2 K. With the applied field increased to 1 T, the normal state is fully restored; the measured upper critical field is consistent with previous results obtained on polycrystalline samples [2]. After subtracting the normal state resistivity data obtained in a field of 1 T from the zero-field data, the onset $T_c$ is identified to 4.5 K as shown in the inset of Figure 2. Below the onset temperature 4.5 K, the resistivity drops rather continuously, although more rapidly than in polycrystalline samples [2]. Despite of the broad transition, zero-field resistivity displays two distinct steps upon decreasing temperature. The resistivity is reduced by more than 80% within the first stage from 4.5 K to ~ 2 K. Below 1.7 K, the resistivity decreases slower than in the first stage and finally vanishes at 1.2 K. This behavior is not due to a poor sample quality, as can be ruled out from our detailed high resolution X-ray diffraction studies, which even allowed the reconstruction of the charge density distributions in $Sc_3CoC_4$ on a sub-atomic scale. [1,13,14]. Thus, the peculiar resistivity behavior of $Sc_3CoC_4$ provides rather a fingerprint of a dimensional crossover: This kind of two-steps transition in resistivity is very similar to $Tl_2Mo_6Se_6$ [7] and has been furthermore frequently observed in proximity junction arrays [30-33]. It has been demonstrated that the initial resistivity drop below the onset $T_c$ originates from the quasi-1D fluctuations in the individual chains, which are in our sample represented by the $[CoC_4]_\infty$ ribbons. Around 2 K, the coupling between neighboring $[CoC_4]_\infty$ ribbons via the Josephson or proximity effect become significant, which results in the second resistivity drop. This coupling effect between individual $[CoC_4]_\infty$ ribbons is also observed in the differential resistivity measurement and will be discussed later. As a result, a true bulk superconducting state with zero resistance, triggered by the transversal Josephson coupling, is achieved below 1.2 K.

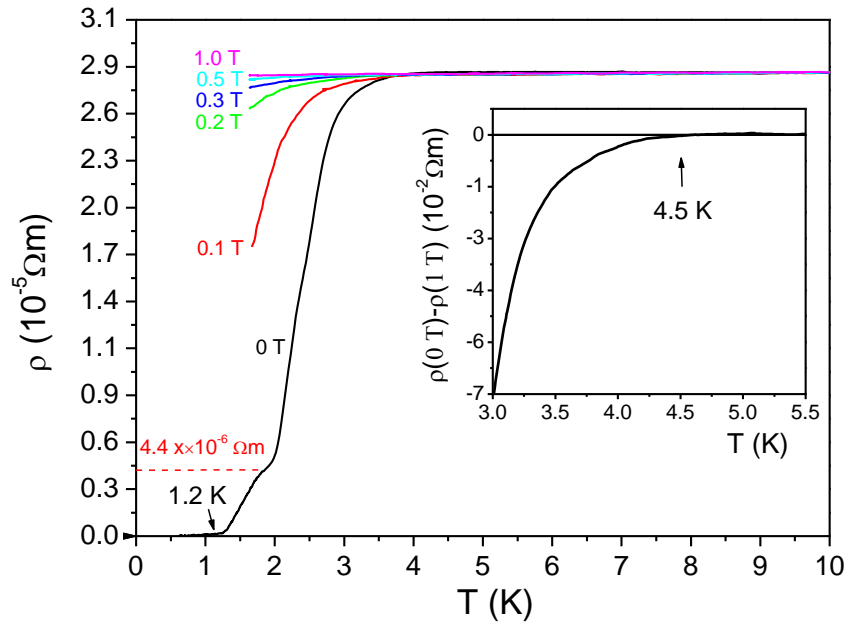

**Figure 2.** Temperature dependence of the electrical resistivity of $Sc_3CoC_4$ in various magnetic fields. The red dashed line marks the resistivity at the onset of the second resistivity drop. Inset: Enlarged view of zero field resistivity after the subtraction of 1 T data, which indicates $T_c^{Onset} = 4.5$ K.

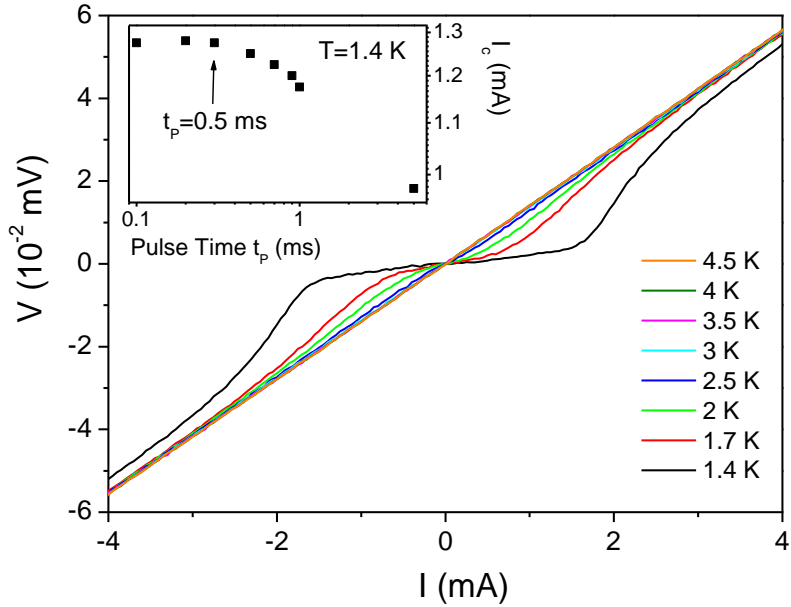

**Figure 3.** Current-Voltage characteristic of $Sc_3CoC_4$ at different fixed temperature with pulse time 0.1 ms. Inset: Critical current as function of pulse time width measured at 1.4 K.

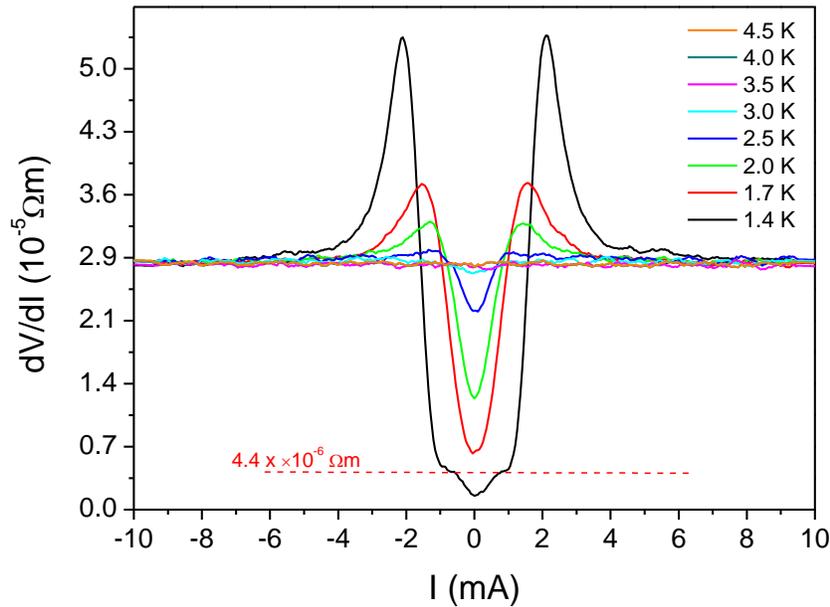

**Figure 4.** Differential resistivity as a function of injected current at different fixed temperatures. The red dashed line indicates the plateau differential resistivity between 0.8 mA and 1 mA.

In Figure 3 we present data of the current-voltage characteristics at various fixed temperatures. As mentioned in the experimental part, the sample is in good thermal contact with a copper block and the current was applied in a very short pulse mode to prevent heating effects. The inset in Figure 3 shows the variation of the critical current with different pulse time widths $t_P$ at 1.4 K. It is obvious that when $t_P$ is kept smaller than 0.5 ms, the critical current is independent of the pulse time width, indicating that the heating effect is negligible. The *I-V* curves show the characteristic ohmic behavior of a normal metallic state above 4.5 K. At lower temperatures, non-linear *I-V* characteristics gradually occur.

Based on the *I-V* data, the differential resistivity as a function of applied current at different temperature shows the continuous development of a supercurrent gap more clearly (Figure 4). he supercurrent gap is barely seen above 3.5 K, although the resistance already dropped by 10 % at this temperature. At 3 K a supercurrent pseudogap is formed and becomes more and more pronounced with decreasing temperature. At 2.5 K, the supercurrent gap forms the characteristic triangular shape of a 1D superconductor, which is dominated by longitudinal phase slips of the order parameter [28,29]. With further decreasing temperature, the gap grows and develops a more rounded form. A tiny second gap with its center approaching zero resistance emerges out at 1.7 K and becomes very pronounced when the temperature reaches 1.4 K. The resistivity plateau at 1.4 K in the range from 0.8 mA and 1 mA, as marked by the red dash line, has a value of $4.4\times10^{-6}$ Ωm. This value is identical to the resistivity in the zero-field $\rho(T)$ curve above the second resistance drop at 2 K (Figure 2). Such intermediate plateaus in d$V$/d$I$ with finite resistance have been also observed in other quasi-1D superconductors [7,10] and it has been demonstrated experimentally and theoretically [34] that they are characteristic fingerprints for a dimensional crossover from a quasi-1D fluctuating state at higher temperature to a 3D bulk phase-coherent state in the low-temperature regime. The intermediate plateau structure at 0.8 mA thus likely originates from local phase-coherent regions triggered by the onset of transversal coupling of the [CoC$_4$]$_\infty$ chains via the Josephson or proximity effect between the [CoC$_4$]$_\infty$ ribbons, whereas the second gap is attributed to a 3D macroscopically phase-coherent superconducting state. Once this transversal coupling is formed, the dimensionality crosses over to 3D, which reduces phase slips in the current path and thus the resistance can drop to zero forming the second gap at small current densities. Larger current densities cause phase slips along the current path and restore a finite differential resistivity value, which forms the $4.4\times10^{-6}$ Ωm plateau.

**Discussion**

Our electrical transport data demonstrates that Sc$_3$CoC$_4$ is a novel quasi-1D superconductor, as suggested by the quasi-1D nature of the [CoC$_4$]$_\infty$ ribbons responsible for superconductivity [1,2]. The material undergoes a complex dimensional crossover within the superconducting transition, which appears in at least 2 stages. It shows a crossover from a quasi-1D fluctuating state in the vicinity of the onset temperature to a 3D global phase coherent superconducting state at very low temperature. This behavior is very similar to other quasi-1D superconductors that we have studied before [7-12], especially the quasi-1D superconductor Tl$_2$Mo$_6$Se$_6$ [7]. The crystalline structure of Tl$_2$Mo$_6$Se$_6$ is similar to Sc$_3$CoC$_4$, where the chains are formed by Mo$_6$S$_6$ clusters, which are separated by loosely-bound Tl$^+$ ions. The superconducting transition of Tl$_2$Mo$_6$Se$_6$ also shows two distinct resistivity drops in the $\rho(T)$ curve and a complex supercurrent gap structure: First a pseudogap of finite resistivity forms and then below a certain temperature an additional bulk supercurrent gap starts to grow out from the center of the pseudogap [7]. In a very similar manner, quasi-1D superconducting fluctuations begin to form in Sc$_3$CoC$_4$ within individual [CoC$_4$]$_\infty$ ribbons below 4.5 K. Consequently, the resistivity starts to deviate from the normal state value and decreases continuously upon cooling. According to the 1D phase-slips model [28,29], the resistivity should depend linearly on the applied current. This explains the pronounced characteristic triangular supercurrent gap in the differential resistivity at 2.5 K, which is reproducibly found in various single crystals and also polycrystalline samples. At this stage, resistivity decreases very continuously, and, above 3 K, hardly any pseudogap is visible in the differential resistivity. This indicates that the coupling between the $[CoC_4]_\infty$ ribbons is negligible [14]. Only when the temperature is decreased below 2 K, a transversal coupling between neighboring $[CoC_4]_\infty$ ribbons begins to become important, which is mediated by either the Josephson or proximity interaction. The second resistivity drop below 2 K and the additional supercurrent gap at 1.7 K are thus fingerprints of this transversal coupling of the chains. The transverse coupling could instantly stabilize phase coherence in the lateral direction and along the ribbons and thus establishes a 3D phase coherent bulk superconducting state with zero resistivity [34]. The specific-heat data presented in Ref. [2] support this scenario: instead of a conventional BCS transition, $C_p$ shows the characteristic broad bump of a quasi-1D superconducting transition [6,9,12] with a maximum occurring slightly above 2.5 K. A crossover towards a 3D scenario is furthermore corroborated by recent flux relaxation measurements, where a change from a single vortex flux creep to a vortex bundle flux creep just below 2.5 K is observed [35]. In absence of this transversal coupling, the resistivity governed by quasi-1D fluctuations would remain finite at all temperatures above 0 K [28,29].

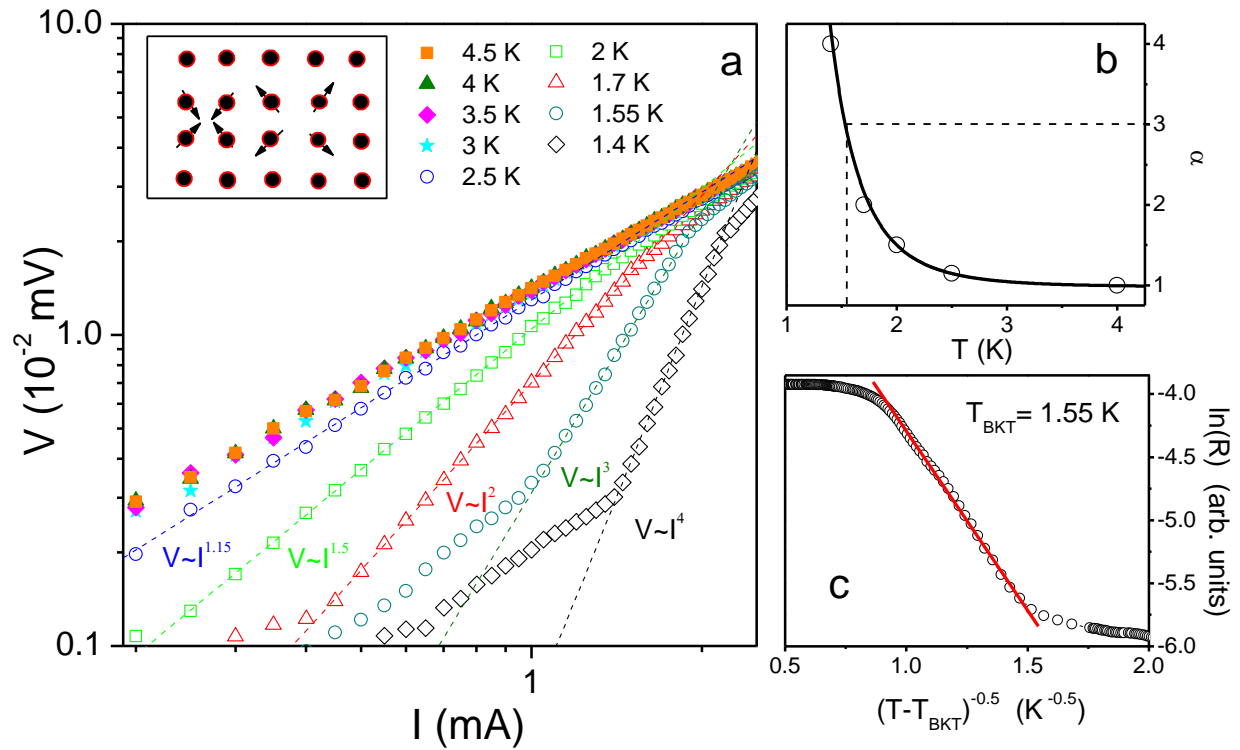

**Figure 5.** **a** Analysis of the *V-I* data in the framework of a Berezinskii-Kosterlitz-Thouless transition in the lateral plane to the $[CoC_4]_\infty$ ribbons. The straight lines in the double logarithmic plot **a** represent power-law fits $V \sim I^{\alpha(T)}$. The inset represents a cartoon to illustrate how the individual phases of the superconducting order parameters in neighboring ribbons can form vortex (right, arrows pointing outwards) and antivortex (left, arrows pointing inwards) excitations in the lateral plane. The ribbons piercing through the lateral plane are illustrated by the dots. **b** Interpolated temperature dependence of the exponent $\alpha$ (which reaches $\alpha = 3$ at $T_{BKT} = 1.55$ K). **c** Expected scaling of the resistivity as a function of temperature, which provides $T_{BKT} = 1.55$ K in agreement with the *V-I* analysis.

It has been demonstrated for $Tl_2Mo_6Se_6$ [7] and for arrays of parallel superconducting nanowires [10], that this 1D to 3D crossover is established via a Berezinskii-Kosterlitz-Thouless-like [17-19] (BKT) transition in the plane perpendicular to the 1D chain direction [7,10]. In such quasi-1D superconducting materials, each chain is characterized by an individual superconducting phase. In the high-temperature regime, the individual phases are not correlated and have arbitrary values. Therefore, there is no global phase coherence in the lateral plane. The superconductor is thus in the 1D limit with finite resistance. When lowering the temperature, the increasing strength of the transverse Josephson or proximity effect results in an exchange coupling of neighboring chains, which triggers vortex / antivortex excitations in the phases of small groups of adjacent chains (see inset of Fig. 5a for an illustration). Below a characteristic temperature $T_{BKT}$, vortices and antivortices become bound in pairs. This triggers macroscopic phase coherence in this lateral plane and has been shown to instantly suppress phase fluctuations along the chains [34]. A BKT-like 2D phase-ordering process in the lateral plane thus triggers a 1D to 3D crossover in such superconducting materials.

To test whether this BKT-like model is applicable to describe the dimensional crossover in $Sc_3CoC_4$, we analyzed the *IV*-characteristics and the resistive transition according to its theoretical predictions. For a superconductor going through a BKT transition, it is expected that the *IV* characteristics follow power law dependencies with $V \sim I^{\alpha(T)}$. For an infinite-size homogenous material, the exponent would be expected to jump from $\alpha = 1$ in the ohmic normal-state to $\alpha = 3$. In a finite-size sample, $\alpha(T)$ rather increases continuously from $\alpha = 1$ to much larger values at lower temperatures [36]. At the characteristic temperature $T_{BKT}$, it is then expected that $\alpha = 3$. In Fig. 5a we plot the *IV* curves on a double logarithmic scale, together with $V \sim I^{\alpha(T)}$ fits. The data follow the predictions well and $\alpha = 3$ occurs at 1.55 K. Note, that the deviations from the power-law dependence in the small current limit at the lowest temperatures are related to the finite size effects [36], which cause a transition to ohmic resistance in the low-current limit. From an interpolation of $\alpha(T)$ (Fig. 5b), we obtain $T_{BKT} = 1.55 \ (\pm 0.05)$ K. Due to the finite size effects, this power law dependence usually occurs over a fairly small range of the *IV* characteristics in the vicinity of the edge of the supercurrent gap. Therefore, as a check of consistency, it is essential to derive $T_{BKT}$ also according to at least a second criterion. A second method of identifying a BKT-like transition is given by the unique temperature scaling of the resistivity above $T_{BKT}$, which originates from thermal activation of the unbound vortex and antivortex pairs. The resistance is expected to vary as $R = A \exp(b/(T_{BKT}-1)^{-1/2})$, with A and b as material constants. This is shown in Fig. 5c. For $T_{BKT} = 1.55 \ (\pm 0.1)$ we find indeed the expected linear relation above $T_{BKT}$, which extends almost over the entire range of the resistive transition, whereas for only slightly smaller or larger values a curvature develops. The scaling behavior predicted by the BKT theory can thus be verified. This suggests that the superconducting transition of $Sc_3CoC_4$ belongs together with $Tl_2Mo_6Se_6$ [7] and superconducting 4 Angstrom carbon nanotube arrays [10] into a new universality class of a quasi-1D superconducting transition, in which a BKT-like transition in the lateral plane triggers a crossover from a 1D fluctuating superconducting state at high temperatures to a 3D bulk superconducting state in the low-temperature regime.

## Conclusions

In summary, we have studied the electrical resistivity, *I-V* characteristics and the supercurrent gap in the differential resistivity of $Sc_3CoC_4$. Quasi-1D superconducting fluctuations were observed below an onset temperature of ~4.5 K. A crossover from quasi-1D superconducting fluctuations to a bulk 3D phase-coherent superconducting state, triggered by a Berezinskii-Kosterlitz-Thouless-like transition in the lateral plane, was discovered by the observation of a second resistivity drop along with an additional supercurrent gap structure below 1.7 K. In comparison with the most prominent example of the quasi-1D superconductor $Tl_2Mo_6Se_6$, $Sc_3CoC_4$ shows a similar complex superconducting transition behavior in form of a dimensional crossover, but with even more pronounced 1D characteristic.

## Acknowledgements

We thank U. Lampe for technical support. This work was supported by the Research Grants Council of Hong Kong Grants SEG_HKUST03, 603010, SRFI11SC02 and the Deutsche Forschungsgemeinschaft (SPP1178).